\documentclass[11pt,a4paper]{article}
\usepackage{jheppub}
\usepackage[T1]{fontenc}
\usepackage{slashed}
\usepackage{physics}
\usepackage[compat=1.0.0]{tikz-feynman}
\usepackage{subcaption}
\usepackage{parskip}
\usepackage[export]{adjustbox}
\usepackage{graphicx}

\newcommand{\mom}{\mathrm{RI'\mbox{-}MOM}}
\newcommand{\smom}{\mathrm{RI\mbox{-}SMOM}}
\newcommand{\momb}{\mathrm{\overline{RI}\mbox{-}MOM}}
\newcommand{\smomb}{\mathrm{\overline{RI}\mbox{-}SMOM}}

\title{Semileptonic weak Hamiltonian to $\mathcal{O}(\alpha \alpha_s)$ in momentum-space subtraction schemes}

\author[a]{M. Gorbahn,}
\author[b]{S. Jäger,}
\author[a]{F. Moretti}
\author[b]{and E. van der Merwe}

\affiliation[a]{Department of Mathematical Sciences, University of Liverpool, Liverpool L69 3BX, United Kingdom}
\affiliation[b]{Department of Physics and Astronomy, University of Sussex, Falmer, Brighton BN1 9QH, United Kingdom}

\emailAdd{Martin.Gorbahn@liverpool.ac.uk}
\emailAdd{s.jaeger@sussex.ac.uk}
\emailAdd{Francesco.Moretti@liverpool.ac.uk}
\emailAdd{ev95@sussex.ac.uk}

\abstract{
  The CKM unitarity precision test of the Standard Model requires a systematic treatment of electromagnetic and strong corrections for semi-leptonic decays.
  Electromagnetic corrections require the renormalization of a semileptonic four-fermion operator. 
  In this work we calculate the $\mathcal{O}(\alpha\alpha_s)$ perturbative scheme conversion between the $\overline{\rm MS}$ scheme and several momentum-space subtraction schemes,
which can also be implemented on the lattice.
We consider schemes defined by MOM and SMOM kinematics and emphasize the importance of the choice of projector for each case.
The conventional projector, that has been used in the literature for MOM kinematics, generates QCD corrections to the conversion factor that do not vanish for $\alpha=0$ and which generate an artificial dependence on the lattice matching scale that would only disappear after summing all orders of perturbation theory. This can be traced to the violation of a Ward identity that holds in the $\alpha =0$ limit.
We show how to remedy this by judicious choices of projector, and define two new schemes $\momb$ and $\smomb$.
We prove that the Wilson coefficients in the new schemes are free from pure QCD contributions, and find that the Wilson coefficients (and operator matrix elements) have greatly reduced scale dependence. Our choice of the $\overline{\rm MS}$ scheme over the traditional \text{$W$-mass} scheme is motivated by the fact that, besides being more tractable at higher orders, unlike the latter it allows for a transparent separation of scales. We exploit this to obtain renormalization-group-improved leading-log and next-to-leading-log strong corrections to the electromagnetic contributions and study the (QED-induced) dependence on the lattice matching scale.
}

\begin{document} 
\maketitle

\section{Introduction}
\label{sec:introduction}
Leptonic and semi-leptonic decays of mesons and nuclear beta decays probe the CKM matrix and provide an electroweak precision test of the standard model (SM), see for example \cite{Cirigliano:2009wk} and \cite{Eric:Crivellin:2020lzu}.
The short-distance physics of meson and nuclear beta decays in the SM is described, to an excellent approximation, by an effective Hamiltonian that involves only a single charged-current operator
\begin{equation}
\label{Effective-semileptonic-operator}
  \mathcal{H}(x)=4\;\frac{G_F}{\sqrt{2}}\;V_{ud}^*\;O(x),\quad  O(x)= \left( \bar{d}(x)\gamma^\mu P_L u(x) \right) \left( \bar{\nu}_l(x)\gamma_\mu P_L l(x) \right),
\end{equation}
where $P_{L}=(1-\gamma^5)/2$ and $G_F$ is the Fermi constant.
At tree-level, the respective Wilson coefficient is directly proportional to the $G_F$ and a single CKM matrix element, here $V_{ud}$.
In particular, the measurements of Kaon \cite{Eric:NA62} and nuclear beta decays \cite{Eric:NuclearReview} test CKM unitarity,  $|V_{ud}^2| + |V_{us}^2| = 1 - |V_{ub}^2|$.
The extraction of the CKM matrix elements relies on the precise predictions of short distance QED and electroweak corrections, a determination of the relevant decay constants and form factors from lattice QCD \cite{Lubicz:2009ht,Bazavov:2012cd,RBCUKQCD:2015joy,Carrasco:2016kpy,FermilabLattice:2018zqv,ETM:2009ptp,RBC:2014ntl,Follana:2007uv,MILC:2010hzw,Durr:2010hr,Durr:2016ulb,QCDSF-UKQCD:2016rau,Dowdall:2013rya,Carrasco:2014poa,Miller:2020xhy,Bazavov:2017lyh}  and the treatment of isospin breaking corrections and long distance QED effects using a combination of chiral perturbation theory and lattice field theory.

Traditionally, the calculation of the short distance contribution relies on current algebra and is performed in the \text{$W$-mass} renormalization scheme \cite{Sirlin:1981ie}.
This scheme preserves the QED Ward identity and ensures that all weak corrections to the Fermi decay can be absorbed into $G_F$ while the short distance corrections for the semi-leptonic decays comprise a large electromagnetic logarithm and electroweak corrections that are mostly absorbed into $G_F$.
QED corrections for leptonic and semi-leptonic decays were calculated both in the current algebra approach \cite{Sirlin:1977sv}, in chiral perturbation theory (\cite{Eric:KnechtChiPT1,Eric:KnechtChiPT2,Eric:NeufeldChiPT}) or in a combined approach with chiral perturbation theory \cite{Eric:Seng19} where the electroweak box diagrams are calculated with lattice gauge theory \cite{Eric:Seng20,Eric:Feng20}.
Another possible scheme is the $\overline{\mathrm{MS}}$ scheme, which is already used for the calculation of QED corrections \cite{vanRitbergen:1999fi,Steinhauser:1999bx} to the Fermi theory that determine $G_F$ as defined in Ref.~\cite{Workman:2022ynf}.
This scheme is also used for the calculation of electroweak corrections to the weak effective Hamiltonian \cite{Gambino:2001au}, where the electroweak matching corrections and next-to-leading order anomalous dimensions for the operator $O$ are given in Ref.~\cite{Brod:2008ss}.
In the $\overline{\mathrm{MS}}$ scheme weak and hadronic scales are separated unlike in the \text{$W$-mass} scheme, and this scale separation simplifies the new physics interpretation and allows for a systematic inclusion of higher-order perturbative corrections.

The complete treatment of QED corrections on the lattice is a difficult task and has so far been performed for purely leptonic decays \cite{Carrasco:2015xwa,Giusti:2017dwk,DiCarlo:2019thl,Boyle:2019rdx}. A novel feature in the semi-leptonic decay is that the relevant operator renormalizes
in the presence of QED corrections. Both the
\text{$W$-mass} scheme and the $\overline{\mathrm{MS}}$ scheme are defined perturbatively and cannot be implemented
on the lattice. This limitation does not apply to momentum-space subtraction schemes, which can be implemented both
on the lattice and in continuum perturbation theory. The renormalization in the RI'-MOM scheme with a lattice
regulator was given in Ref.~\cite{DiCarlo:2019thl}, including the one-loop perturbative matching to the \text{$W$-mass} scheme.

In this paper we will perform the perturbative matching at two-loop level for different momentum-space subtraction  schemes.  These schemes are regulator-independent (RI) and are defined through a condition on a projected renormalized Green's function for a particular off-shell momentum configuration.
The choice of projector is an important part of the definition of a particular RI scheme. In particular,
special choices of projectors are required to
ensure that the weak currents do not receive a finite renormaliation in RI schemes \cite{Gracey_2003,Gracey_2011}.
Similarly, it is preferable to choose renormalization conditions that do not result in a finite renormalization of the
semileptonic operator $O$ in the pure QCD limit, as a finite QCD renormalization would imply an artificial (residual)
scale dependence that only (formally) disappears once all orders of perturbation theory are summed.

The remainder of this paper is organised as follows:
In Section \ref{sec:renorm-cond} we discuss different choices of renormalization scheme and show which choices of projectors lead to vanishing pure-QCD corrections.
Section \ref{sec:details-calculation} describes salient technical aspects of our two-loop calculation, including the tensor reduction and the master integrals used in the loop calculations.
In Section \ref{sec:results--numerics} we present the results, where we combine the two-loop lattice continuum matching corrections with the known short-distance corrections and perform a renormalization-group improvement showing
explicitly the dependence on the scales $\mu_W$, $\mu_b$, and the lattice matching scale $\mu_L$.
We also study in detail the cancellation of the dependence on the scale $\mu_L$ between the
RG-improved Wilson coefficient and the conversion factor (or matrix element).
Section \ref{sec:conclusions} contains our conclusion.

\section{Renormalization conditions and change of scheme}
\label{sec:renorm-cond}

\begin{figure}
  \centering
  \begin{tabular}{cc}
    \includegraphics[width=0.4\textwidth]{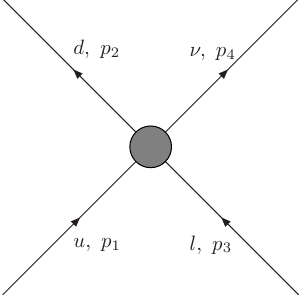}&
    \includegraphics[width=0.4\textwidth]{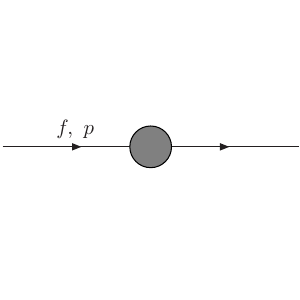}
  \end{tabular}
\caption{Kinematic conventions for the four- and two-point diagrams.} 
\label{fig:off-shell-kinematics}
\end{figure}

In this section we define the $\overline{\mathrm{MS}}$, $\mom$, $\smom$, $\momb$, and $\smomb$ renormalization
schemes for the fields and the operator $O$ and express the scheme conversion factors in terms of two- and four
point Green's functions (Figure \ref{fig:off-shell-kinematics}). 

Let us define the connected fermion two-point function $S$ and the amputated four-point
function with $O$-insertion $\Lambda$ through (respectively)
\begin{equation}
\label{fermion propagator}
(2\pi)^4 iS(p)_{\alpha \beta}\delta^4(p-q)\delta^{ij}=\int d^4 x d^4 y e^{i(p\cdot x - q\cdot y)}\langle 0|T\{\psi^{i}_{\alpha}(x) \overline{\psi^{j}_\beta}(y)\}|0\rangle
\end{equation} 
and
\begin{equation}
\label{amputated green's function}
\begin{split}
&\int d^4x d^4x_{1,...,4} e^{-i(p\cdot x +p_1\cdot x_1 -p_2\cdot x_2+p_3\cdot x_3 -p_4\cdot x_4)} \langle
0|T\{\overline{\psi^{u,i}}_\beta(x_1)\psi^{d,j}_\alpha(x_2)\overline{\psi^l}_\delta(x_3)\psi^{\nu}_{\gamma}(x_4)O(x)\}|0\rangle=\\
&(2\pi)^4 \delta^{ij} S(p_1)_{ \beta^{\prime} \beta} S(p_2)_{\alpha \alpha^{\prime}}S(p_3)_{\delta^{\prime} \delta}S(p_4)_{\gamma \gamma^{\prime}}\Lambda_{O}(p_1,p_2,p_3,p_4)_{\alpha^{'} \beta^{'} \gamma^{'} \delta^{'}}\delta^4(p-p_1+p_2-p_3+p_4) ,
\end{split}
\end{equation} 
where ${i,j}$ represent colour indices, which are absent in \eqref{fermion propagator} in the case of leptons\footnote{ In the original version of this work, Eq.~\eqref{amputated green's function} had the indices contracted incorrectly.}.
We recall that the 1PI two-point function is then given by $S^{-1}$.
Fully defining the fields $\psi_f$ (f = u, d, $\ell$) and the operator $O$ requires renormalization conditions.
We note that, because $O$
does not mix with other operators, any two schemes $A$, $B$ differ only by a (finite) rescaling:
\begin{eqnarray}
   O^A &=& {\cal C}_O^{A \to B} O^B,   \label{eq:COdef}   \\
   \psi_f^A &=& ({\cal C}_f^{A \to B})^{1/2} \psi_f^B,   \label{eq:Cpsidef}
\end{eqnarray}
from which the relation $C_O^B = {\cal C}_O^{A \to B} C_O^A$ for the Wilson coefficient follows.

The $\mom$, $\smom$, $\momb$, and $\smomb$ schemes are defined by imposing the renormalization conditions
\begin{equation}
\label{eq:RIcond2}
\sigma^A \equiv \frac{1}{4\;p^2}{\rm Tr}\left(({S^A})^{-1}(p)\slashed{p}\right) \stackrel{\mbox{\small A=RI}}{=} 1
\end{equation}
and
\begin{equation}
\label{eq:RIcond4}
\lambda^A \equiv \Lambda_{\alpha \beta \gamma \delta}^A {\mathcal{P}^{\beta \gamma \delta \alpha}}
\stackrel{\mbox{\small A=RI}}{=}1
\end{equation}
at suitable kinematics, where ${\cal P}$ is a constant Dirac tensor satisfying
$\Lambda^{\rm (tree)}_{\alpha \beta \gamma \delta} \, {\mathcal{P}^{\beta \gamma \delta \alpha} }=1$.\footnote{In the original version of this work, Eq.~(\ref{eq:RIcond4}) had the indices contracted incorrectly. }
The difference between $\mom$ and $\momb$ (as well as $\smom$ and $\smomb$) is only in the choice of projector, with the latter schemes adding conditions to specify the projector which are not present in the former. We defer specifics to Section \ref{sec:RIspecifics} below.

It follows that the scheme conversion factors satisfy
\begin{eqnarray}
{\cal C}_f^{A \to \rm RI} &=& \left( \sigma^A \right)^{-1/2} ,  \label{eq:Cfmaster}  \\
{\cal C}_O^{A \to \rm RI} &=& \lambda^A \left( \sigma^A_u \sigma^A_d \sigma^A_\ell \right)^{1/2} , \label{eq:COmaster}
\end{eqnarray}
with implicit dependence on the choice of kinematic point and projector.

Equations (\ref{eq:Cfmaster}) and (\ref{eq:COmaster}) are the master formulas allowing for the computation of
the scheme conversion factors. We emphasize that they hold independently of the choice of regulator
used in computing the two- and four-point functions, though defining the {$\overline{\rm MS}$} scheme in practice
entails the use of dimensional regularization.

\subsection{Dimensional regularization and $\overline{\rm MS}$ renormalization}\label{subsec:MSrenormalization}

In our computations we employ dimensional regularization (with an anticommuting $\gamma^5$; no ambiguous traces
will occur in the following), which at the same time is the basis for defining
$\overline {\rm MS}$ schemes.  The relations between the ${\overline{\rm MS}}$-renormalized
objects $O^{\overline{\rm MS}}$ and
$\psi^{\overline{\rm MS}}$ and the bare objects $O^b$, $\psi^b$ in dimensional regularization are analogous to
(\ref{eq:COdef}), (\ref{eq:Cpsidef}) but complicated by a regularization artefact, the evanescent operators.
These are chosen such that their (renormalized) infrared-finite Green's functions vanish as $d \to 4$; in particular
they will vanish at the RI subtraction point. In our case and to the loop order of our calculation, a single evanescent
operator suffices, the bare version of which can be chosen to be

\begin{equation}
\label{evanescent operator}
E = (\bar{d} \gamma^\mu \gamma^\nu \gamma^\lambda P_L u) (\bar{\nu}_l \gamma_\mu \gamma_\nu \gamma_\lambda P_L \ell) - (16-4\epsilon-4\epsilon^2) (\bar d \gamma^\mu P_L u) (\bar \nu_{\ell} \gamma_\mu P_L \ell),
\end{equation} 
following the notation of \cite{Gorbahn:2004my} and where all fields are understood to be bare.

We define 
\begin{align}
 \psi^b_f &= \big(Z_{2,f}^{\overline{\mathrm{MS}}}\big)^{1/2}\psi^{\overline{\rm MS}}_f \\
    O^{\overline{\mathrm{MS}}} &=Z^{\overline{\mathrm{MS}}}_{OO}\; O^b+ Z^{\overline{\mathrm{MS}}}_{OE}\; E^b,
\end{align}
from which it follows that
\begin{align}
    \Lambda_O^{\overline{\mathrm{MS}}} &=
  \big(Z_{2,u}^{\overline{\mathrm{MS}}}\big)^{1/2}\big(Z_{2,d}^{\overline{\mathrm{MS}}}\big)^{1/2}\big(Z_{2,l}^{\overline{\mathrm{MS}}}\big)^{1/2}
  \left(Z^{\overline{\mathrm{MS}}}_{OO}\;\Lambda_O^b + Z^{\overline{\mathrm{MS}}}_{OE}\;\Lambda_{E}^b \right), \\
S^{\overline{\mathrm{MS}}} &= \big(Z^{\;\overline{\mathrm{MS}}}_{2,f}\big)^{-1} S^b.
\end{align}

Explicitly,
\begin{align}
&Z_{2,f}^{\overline{\mathrm{MS}}}=1-\frac{\alpha}{4\;\pi}\;\frac{\xi\;q^2}{\epsilon}-\frac{\alpha_s}{4\;\pi}\;\frac{C_F\;\xi_s}{\epsilon }+\frac{\alpha}{4\;\pi}\;\frac{\alpha_s}{4\;\pi}\;\left(\frac{C_F\;\xi\;\xi_s\; q^2}{\epsilon
   ^2}+\frac{3\;C_F\;q^2}{2\;\epsilon }\right), \label{Z2 MSbar}\\
&Z^{\overline{\mathrm{MS}}}_{OO}=1-\frac{\alpha}{4\;\pi}\;\frac{2}{\epsilon}+\frac{\alpha}{4\;\pi}\;\frac{\alpha_s}{4\;\pi}\;\left(\frac{7\;C_F}{4\;\epsilon}\right),\label{ZOO MSbar}\\
&Z^{\overline{\mathrm{MS}}}_{OE}=-\frac{\alpha}{4\;\pi}\;\frac{1}{12\;\epsilon } +  O\left(\alpha\;\alpha_s\right) \,,
\label{ZOE MSbar}
\end{align}
where $\alpha$ and $\alpha_s$ represent the electromagnetic and strong coupling constant respectively, while $\xi$ and $\xi_s$ represent the photon and gluon gauge fixing parameter. The objects $\Lambda_O^{\overline{\mathrm{MS}}}$ and $S^{\overline{\mathrm{MS}}}$ then determine
${\cal C}^{\overline{\mathrm{MS}} \to \rm RI}$ as previously described.

\subsection{Specifics of the RI schemes and Ward identity}
\label{sec:RIspecifics}
The main aim of the present paper is to compute the conversion factor between the $\overline{\mathrm{MS}}$ scheme (as
defined above) and improved versions of two momentum-space subtraction schemes defined in the literature:
\begin{itemize}
\item
$\mom$ \cite{Martinelli:1994ty};
\item
$\smom$ \cite{Sturm_2009}.
\end{itemize}

The two schemes are characterised by different kinematics and projectors. In the $\mom$ scheme, all four external momenta in Figure \ref{fig:off-shell-kinematics} are equal, while $\smom$ employs a symmetric configuration with two independent momenta such that
\begin{align}
  \label{RI-MOM config}
\mom:  &\qquad \qquad p_1=p_2=p_3=p_4=p, \quad p^2=-\mu^2,\\
\label{RI-SMOM config}
\smom: &\qquad \qquad p_1=p_3,\quad p_2=p_4,\quad p_1^2=p_2^2=-\mu^2, \quad p_1\cdot p_2=-\frac{1}{2}\mu^2.
\end{align} 

In both schemes, the condition (\ref{eq:RIcond2}) is imposed.

The conventional definition of the projector ${\cal P}$ entering the condition (\ref{eq:RIcond4}) on the renormalized four-point function is \cite{Carrasco:2015xwa,ProjectorDefinition1}
\begin{equation}
\label{eq:projstd}
\mathcal{P}^{\alpha \beta \gamma \delta} =
- \frac{1}{16} \left( \gamma^\mu P_R \right)^{\alpha \beta} \left( \gamma_\mu P_R \right)^{\gamma \delta}
\equiv - \frac{1}{16}  \left( \gamma^\mu P_R\otimes \gamma_\mu P_R  \right)^{\alpha\beta\gamma\delta},
\end{equation}
where $P_R=(1+\gamma^5)/2$. (This results in the usual single trace involving both fermion lines for the standard choice of projector \cite{Carrasco:2015xwa}, as well as all projectors considered in this work (see below).)

This choice of the projector for \eqref{Effective-semileptonic-operator} leads to a scale dependence of the semileptonic
operator already in pure QCD. Such a scale dependence does not occur in the $\overline{\rm MS}$ scheme and, as
we explain in the following, its presence in the standard RI schemes can be traced to the violation of a Ward identity
which appears in the pure-QCD limit. Such an artificial scale dependence is undesirable from a conceptual perspective
and complicates error control when perturbative and lattice results are eventually combined. We derive projectors below which preserve the Ward identity and ensures that the running is of order $\alpha_{\rm em}$. Correspondingly, the
conversion factors ${\cal C}^{\overline{\rm MS}\to \overline{\rm RI}}$ between the lattice schemes and the $\overline{{\rm MS}}$
scheme are modified at $O(\alpha_s)$ and $O(\alpha_s^2)$ relative to the conventional projectors. The Ward-identity-preserving $\overline{{\rm RI}}$ projectors are:

\begin{equation}
\label{RI'-MOM projector}
\mathcal{P}^{\momb}=\frac{1}{12\;\mu^2}\Big(\slashed{p} P_R \otimes \slashed{p} P_R - \frac{\mu^2}{2}\gamma^\nu P_R\otimes \gamma_\nu P_R\Big),
\end{equation}
\begin{equation}
\label{RI-SMOM projector}
\begin{split}
\mathcal{P}^{\smomb} =\frac{1}{4}\Big(
&-\frac{1}{2}\gamma^\nu P_R \otimes \gamma_\nu P_R-\frac{1}{\mu^2}\slashed{p}_1 P_R\otimes \slashed{p}_1 P_R-\frac{1}{\mu^2}\slashed{p}_2 P_R\otimes \slashed{p}_2 P_R+\\
&+\frac{1}{\mu^2}\slashed{p}_1 P_R\otimes \slashed{p}_2 P_R+\frac{1}{\mu^2}\slashed{p}_2 P_R\otimes \slashed{p}_1 P_R
\Big).
\end{split}
\end{equation}

To see how these projectors are obtained, first note that, if electromagnetism is neglected, no diagrams with propagators connecting the quark and lepton lines occur. The lepton line just gives the tree-level leptonic current $L_\mu = \gamma_\mu P_L$ (again, we use open indices here). Hence (suppressing Dirac indices)
\begin{equation}  \label{eq:Lambda_fact}
  \Lambda^b = \Lambda^{b,\mu}(p_1, p_2) \otimes \gamma_\mu P_L  + {\cal O}(\alpha) \, ;
\end{equation}
note that, with our choice of Fierz ordering, this applies to the bare Green's functions in dimensional regularization. 

Now, $\Lambda^\mu$ is the 1PI vertex function in pure QCD for the conserved current $j^\mu = \bar d\gamma^\mu P_L u$, which  satisfies the Ward identity
\begin{equation}
    \label{general Ward identity}
    (p_1 - p_2)_\mu \Lambda^{b, \mu}(p_1,p_2)=S^b(\slashed{p}_1)^{-1} - S^b(\slashed{p}_2)^{-1} ,
\end{equation}
which in the exceptional configuration reads
\begin{equation}
\label{Ward identity one momentum}
\Lambda^{b, \mu}(p_1=p_2=p)= \frac{\partial}{\partial p_\mu} S^b(\slashed{p})^{-1}.
\end{equation}
More precisely, these identities hold in dimensional regularization with anticommuting $\gamma^5$,
and continue to hold after minimal subtraction. (As is well known, there exist regularizations in which the Ward identity does \textit{not} hold.) As a consequence,
the current does not renormalize in $\overline{\rm MS}$,
and for the semileptonic operator we have $Z^{\overline{\rm MS}}_{OO} =1+ {\cal O}(\alpha)$.
It follows that the anomalous dimension is ${\cal O}(\alpha)$ and the Wilson coefficient does not run in $\overline{\rm MS}$ in pure QCD.

As Gracey has pointed out in his work on momentum-space subtraction schemes for quark bilinear operators \cite{Gracey_2011}, preserving the Ward identity requires a judicious choice of projector. Unfortunately, (\ref{general Ward identity}, \ref{Ward identity one momentum}) do not hold   (with
QED neglected and $\Lambda^\mu$ defined as above) for the $\mom$ and
$\smom$ projectors when applied to the semileptonic operators, and as a result $Z^{\rm RI}_{OO} =1+ {\cal O}(\alpha_{\rm s}, \alpha)$.
The resulting pure QCD renormalization for these projectors is finite so that the Wilson coefficient still does not run at $\mathcal{O}(\alpha_s)$.
Yet, the resulting scheme conversion factor carries an implicit scale dependence, due to the truncation of 
the perturbation series, which leads to unnecessarily large theoretical uncertainties on the Wilson coefficient (or
operator matrix element).

To find suitable replacements for the conventional projector  (\ref{eq:projstd}), we extend the idea in Ref.~\cite{Gracey_2011} and expand the four-point function in a set of
basis structures $\mathcal{T}_{(k)}(p_1,p_2))$ and Lorentz-invariant form factors $F_k (p_1,p_2)$,
\begin{equation}
    \label{parameterisation of Green's function}
     \Lambda(p_1,p_2)|_{p_1^2=p_2^2=-\mu^2} = \sum_k F_k(p_1,p_2) \mathcal{T}_{(k)}(p_1,p_2) .
\end{equation}
In pure QCD, for general kinematics, the following 6 structures are sufficient:
\begin{equation}
\label{list of QCD final structures}
\begin{split}
            &\mathcal{T}_{(1)}(p_1,p_2)=\gamma^\mu P_L\otimes\gamma_\mu P_L, \\
            &\mathcal{T}_{(2)}(p_1,p_2)=\frac{1}{\mu^2}\slashed{p}_1 P_L \otimes\slashed{p}_1 P_L, \\
            &\mathcal{T}_{(3)}(p_1,p_2)=\frac{1}{\mu^2}\slashed{p}_1 P_L\otimes \slashed{p}_2 P_L, \\
            &\mathcal{T}_{(4)}(p_1,p_2)=\frac{1}{\mu^2}\slashed{p}_2 P_L\otimes \slashed{p}_1 P_L, \\
            &\mathcal{T}_{(5)}(p_1,p_2)=\frac{1}{\mu^2}\slashed{p}_2 P_L\otimes \slashed{p}_2 P_L,\\
            &\mathcal{T}_{(6)}(p_1,p_2)=\frac{1}{\mu^2}\gamma^\mu\slashed{p}_2\slashed{p}_1 P_L \otimes\gamma_\mu P_L .
    \end{split}
\end{equation}

The ordering in $\mathcal{T}_{(6)}$ is such that all structures but the first one vanish by the equations of
motion (but of course not at general or SMOM kinematics).
Note that no evanescent structures occur in the pure-QCD limit.
For MOM kinematics \eqref{RI-MOM config} the shorter basis
\begin{equation}
\label{list of MOM QCD final structures}
\begin{split}
&\mathcal{T}_{(1)}(p_1=p_2=p)=\gamma^\mu P_L \otimes\gamma_\mu P_L, \\
&\mathcal{T}_{(2)}(p_1=p_2=p)=\frac{1}{\mu^2}\slashed{p} P_L \otimes\slashed{p} P_L
\end{split}
\end{equation}
suffices. 

All structures are easily rewritten with $\gamma_\mu P_L$ as the second (``leptonic'') factor, 
e.g.\ $\mathcal{T}_{(2)}=\frac{1}{\mu^2} p_1^\mu \slashed{p}_1 P_L \otimes \gamma_\mu P_L$.
Noting that $S(\slashed{p})^{-1} =\Sigma(p^2) \slashed{p}$
and comparing coefficients of $\slashed{p}_1$ and $\slashed{p}_2$
on both sides of the Ward identity, the form factors must satisfy, for general kinematics,
\begin{equation}
\label{Ward identity conditions SMOM}
    \begin{cases}
    F_1(p_1,p_2) - \frac{1}{2} F_2(p_1,p_2) + \frac{1}{2} F_3(p_1,p_2) = \Sigma(p_1^2) \\
    F_1(p_1,p_2) + \frac{1}{2} F_4(p_1,p_2) - \frac{1}{2} F_5(p_1,p_2) - F_6(p_1,p_2)  = \Sigma(p_2^2), 
    \end{cases}
\end{equation}
\normalsize
which, for $p_1^2 = p_2^2 = - \mu^2$, can be combined into a family of equations 
\begin{equation}
    \label{Ward identity conditions SMOM on final amplitudes}
    \begin{split}
		&x(F_1(p_1,p_2) -\frac{1}{2} F_2(p_1,p_2) + \frac{1}{2} F_3(p_1,p_2))+\\
		&+(1-x)(F_1(p_1,p_2)+\frac{1}{2} F_4(p_1,p_2) - \frac{1}{2} F_5(p_1,p_2) - F_6(p_1,p_2))=\Sigma(-\mu^2).
	\end{split}
\end{equation}
whose left-hand side equates to $\Sigma(-\mu^2)$.
This allows us to define an infinite\footnote{The symmetry $p_1 \leftrightarrow p_2$ at the symmetric point results in additional constraints on the form factors. In particular, it follows that $F_2(p_1,p_2) = F_5(p_1,p_2)$, which has been explicitly checked at two-loop in QCD \cite{Gracey_2011}. One could use this property to further increase the space of possible projectors.} number of potential projectors parameterised by $x$.
For MOM kinematics, the simpler condition
\begin{equation}
\label{Ward identity conditions MOM}
F_1(p_1=p_2=p)=\Sigma(- \mu^2)
\end{equation}
must hold.

As explained above, the conditions (\ref{Ward identity conditions SMOM on final amplitudes}) and (\ref{Ward identity conditions MOM}) are satisfied in the $\overline{\rm MS}$ scheme.
They will hold in a momentum-space subtraction scheme if the projector ${\cal P}$ is defined such that the projected amplitude results in the left-hand sides of
(\ref{Ward identity conditions SMOM on final amplitudes}) (for $\smomb$) and 
(\ref{Ward identity conditions MOM}) (for $\momb$), in which case
${\cal C}^{\overline{\rm MS}\to {\rm \overline{RI}}} = 1$ (in pure QCD) by virtue of (\ref{eq:COmaster}).
This is equivalent to the
current conservation condition $Z_{OO}^{\rm \overline{RI}} =1+ {\cal O}(\alpha)$, and
at the same time shows that the Wilson coefficient
necessarily agrees with $\overline{\rm MS}$ up to corrections suppressed by the electromagnetic coupling
constant. In other words, we require
\begin{eqnarray}
 {\cal P}(\mathcal{T}^\mu_{(i)}) &=&
   \left\{ 1, -\frac{x}{2}, \frac{x}{2}, \frac{1-x}{2}, -\frac{1-x}{2}, x-1 \right\}
       \qquad (\smomb),    \label{eq:SMOMprojcond}
\\
  {\cal P}(\mathcal{T}^\mu_{(i)}) &=&
     \left\{ 1, 0 \right\}
      \qquad \qquad  \qquad \qquad \qquad \qquad \qquad\; (\momb) .     \label{eq:MOMprojcond}
\end{eqnarray}

To find solutions to  (\ref{eq:SMOMprojcond}) and (\ref{eq:MOMprojcond})
 we define a ``basis'' of linearly independent projectors (six for SMOM and two for MOM) as
\begin{equation}
    \label{projectors}
    \mathcal{P}_{(k)}=\mathcal{T}_{(k)}|_{P_L \rightarrow P_R}.
\end{equation}
Eqs.~(\ref{eq:SMOMprojcond}) and (\ref{eq:MOMprojcond}) then provide linear systems which uniquely
determine the projectors in terms of our basis;
the results are given in (\ref{RI'-MOM projector}) and (\ref{RI-SMOM projector}).
(There may exist other suitable projectors built from
different basis structures.) The $\smomb$ projector for general $x$ is
\begin{equation}
\begin{split}
{\cal P}^{\smomb} = \frac{x-2}{12}{\cal P}_{(1)} - \frac{x+1}{6}{\cal P}_{(2)}+\frac{2-x}{6}{\cal P}_{(3)}+\\+\frac{1+x}{6}{\cal P}_{(4)}+\frac{x-2}{6}{\cal P}_{(5)}+\frac{2x-1}{12}{\cal P}_{(6)}
\end{split}
\end{equation}
This projector reduces to the one defined in (\ref{RI-SMOM projector}) for $x=\frac{1}{2}$, which we use as a reference value to present our results.
As the conventional projectors appear among our basis projectors but do not agree with our solutions, it also follows
that the standard schemes do not preserve the Ward identity. This is also evident from the known results for
the Wilson coefficient, which has a scheme conversion factor different from unity, even in the absence of electromagnetism (see Section \ref{sec:results--numerics}). 

\subsection{\text{$W$-mass} renormalization scheme and the definition of the Fermi constant}

The \text{$W$-mass} renormalization scheme \cite{Sirlin:1977sv,Sirlin:1981ie} was traditionally used in the determination of the Fermi constant $G_F$ and is still in use in the calculation of electroweak corrections for the semi-leptonic decays \cite{Sirlin:1981ie,Eric:Seng19,DiCarlo:2019thl}. 
In this scheme, the amplitude is regularized by splitting the photon propagator 
\begin{equation}
\label{W-Mass photon propagator}
\frac{1}{q^2}\longrightarrow \frac{1}{q^2 - M_W^2} + \frac{M_W^2}{M_W^2-q^2}\frac{1}{q^2},
\end{equation}
where $q$ is the momentum carried by the photon and $M_W$ is the mass of the W-Boson. 
The first term of \eqref{W-Mass photon propagator} acts as a massive photon propagator that contains all UV poles,
which are absorbed by $G_F$.
The second term is UV finite, thanks to the $W$-boson mass acting as a hard UV cut-off, but results in an IR contribution
to the Fermi constant of $\mathcal{O}(\alpha m_{\mu}^2/M_W^2)$. When $G_F$ is used to normalize the
weak Hamiltonian, such a contribution, while small, breaks the manifest separation of scales that is a main virtue of the
effective-field-theory approach.

On the other hand the complete 2-loop QED corrections to the Fermi theory (leptonic weak Hamiltonian)
have been calculated in the $\overline{\mathrm{MS}}$ scheme \cite{vanRitbergen:1999fi,Steinhauser:1999bx} for the Fermi operator in its Fierz-rearranged form, and this scheme was used for the determination of $G_F$ from the muon
lifetime in ~\cite{Workman:2022ynf}.
This definition of $G_F$ is also used in the calculation of electroweak corrections to the weak effective Hamiltonian \cite{Gambino:2001au,Brod:2008ss}, where the normalization of the dimension-6 Hamiltonian to $G_F$ absorbs most electroweak corrections. In the present work, we employ the
${\overline{\mathrm{MS}}}$ scheme for $G_F$. Our Wilson coefficient results
below can therefore directly be used with $G_F$ from~\cite{Workman:2022ynf}, and allow for a transparent
separation of contributions from different scales.

In order to be able to compare to the result in Ref.~\cite{DiCarlo:2019thl} 
we have however determined the scheme conversion from the $\overline{\mathrm{MS}}$ scheme to the \text{$W$-mass} scheme to one loop.
Neglecting $\mathcal{O}(\alpha m_{\mu}^2/M_W^2)$ and  $\mathcal{O}(\alpha m_s^2/M_W^2)$ corrections, we find that the conversion factor for the Fermi (leptonic) operator equals one, while the conversion factor for the semi-leptonic operator reads
\begin{equation}
\label{converion MS W-Reg}
\mathcal{C}_O^{W-{\rm mass}\to \overline{\mathrm{MS}}}=1-\frac{\alpha}{4\pi}\frac{11}{3}.
\end{equation}

\section{Details of the calculation}
\label{sec:details-calculation}
\begin{figure}
\centering
\begin{tabular}{ccc}
\includegraphics[width = 0.3\textwidth]{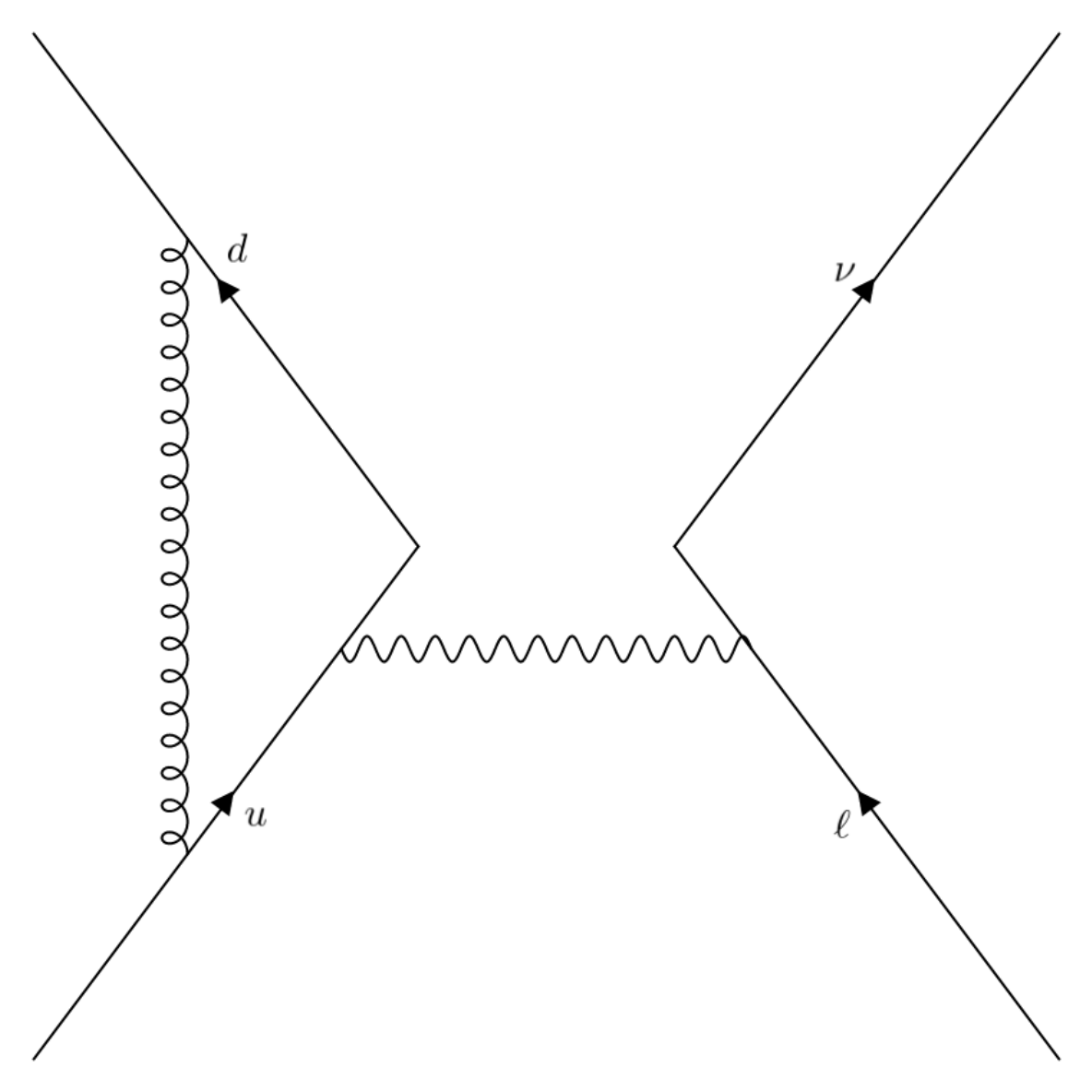}&
\includegraphics[width = 0.3\textwidth]{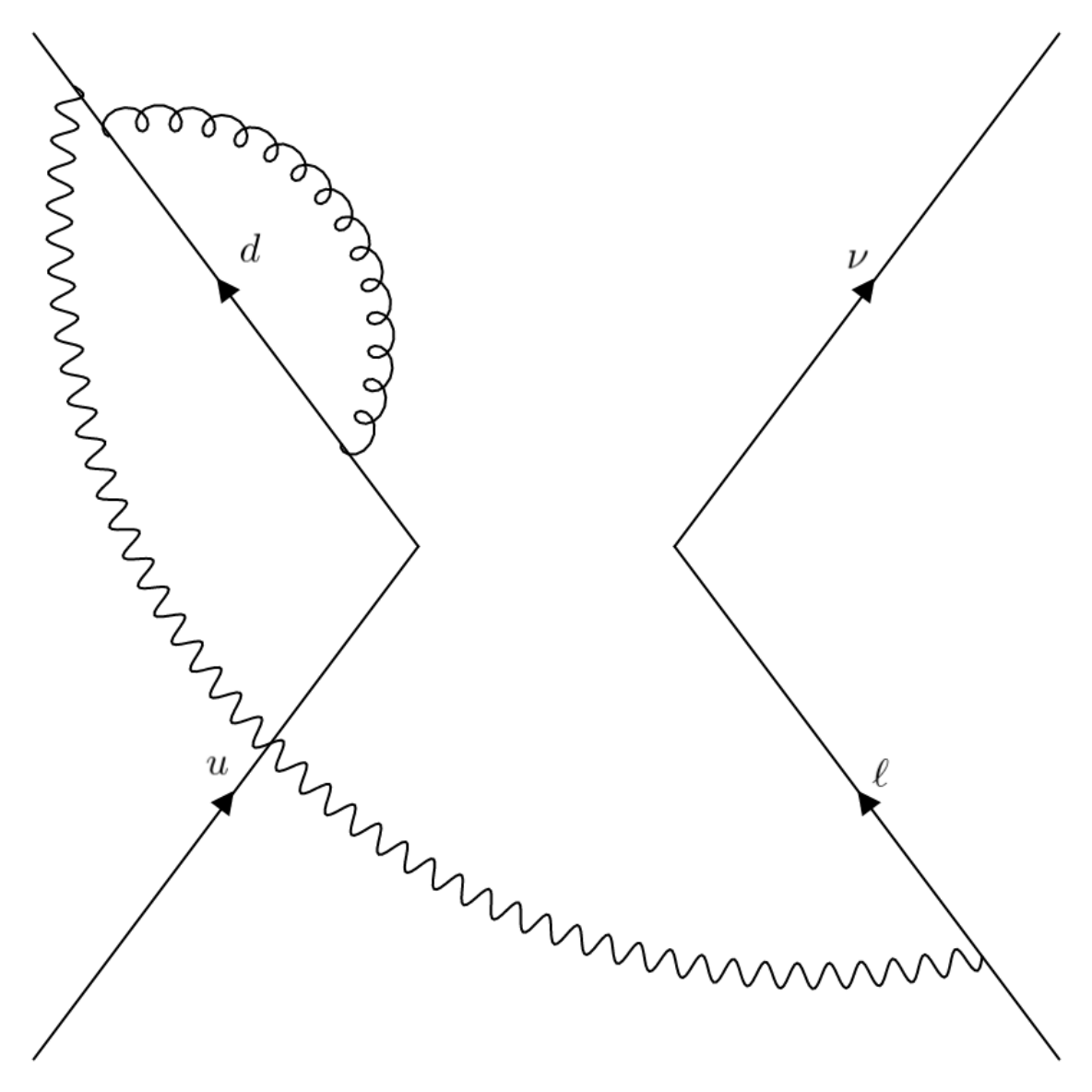}&
\includegraphics[width = 0.3\textwidth]{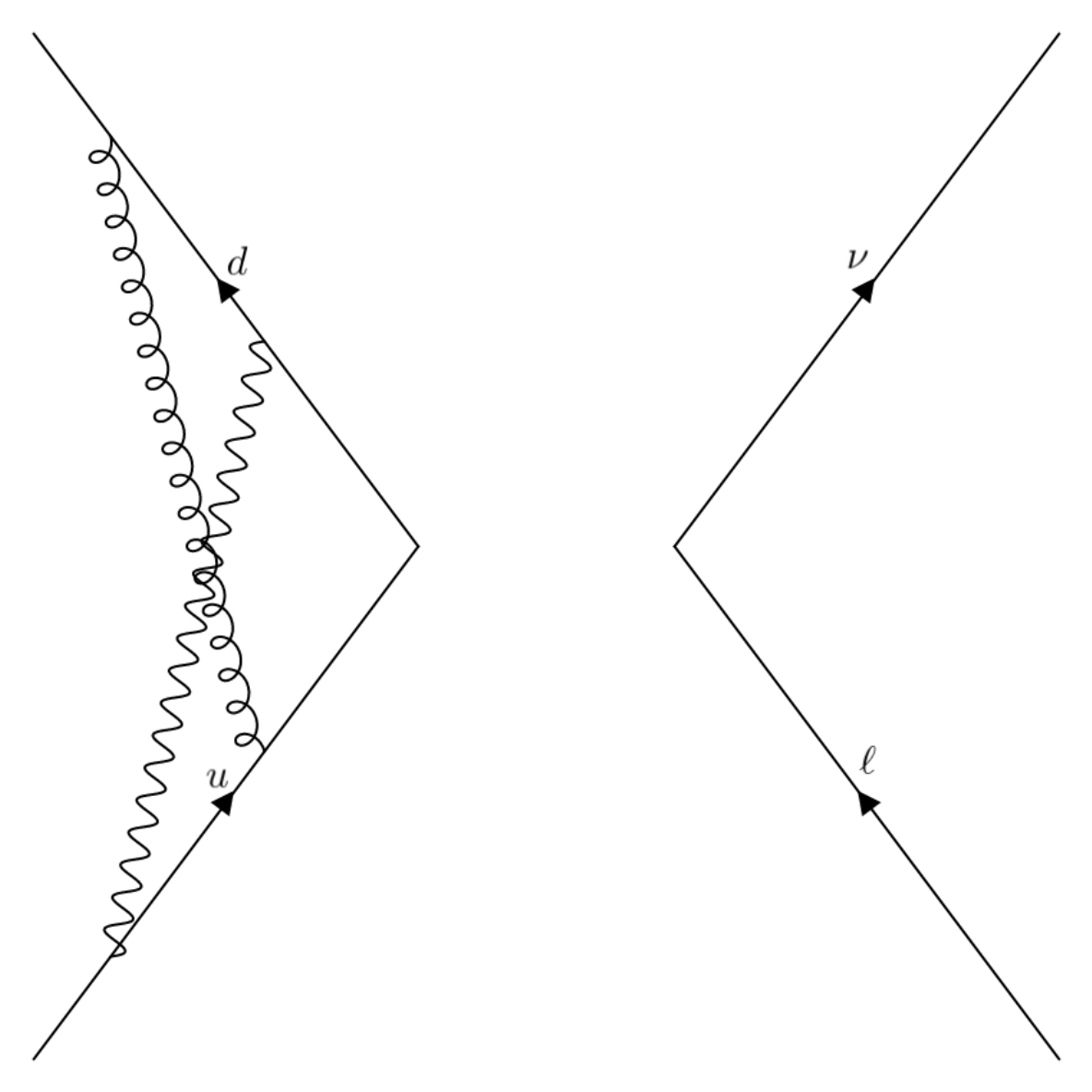}
\end{tabular}
\caption{Some examples of the two-loop diagrams calculated. They are from Basis 1, 2, and 4 respectively.}
\label{examplediagrams}
\end{figure}

There are 21 relevant diagrams at $O(\alpha \alpha_s)$ for the renormalization of semi-leptonic operator and example diagrams can be seen in Figure \ref{examplediagrams}.

We kept open Dirac indices in the evaluation of the respective Feynman amplitudes and only took the traces with the projectors as defined in Section \ref{sec:renorm-cond} after renormalization.
This allows us to consider different projectors and has the additional benefit that possible ambiguities arising from the treatment of gamma matrices, specifically $\gamma^5$, in $d$-dimensions are avoided. 
In our loop calculations tensor integrals appear.

At two-loops, the most complicated structure is given by
\begin{equation}
I^{\mu\nu\rho\sigma}=\int d^dk\;d^dl\; \frac{k^\mu k^\nu l^\rho l^\sigma}{D(p_1,p_2)}\,,
\end{equation}
where $k$ and $l$ are the loop momenta and $D(p_1,p_2)$ is a combination of the propagators involved in the loop.

We applied the Passarino-Veltman decomposition \cite{PASSARINO1979151} to write the tensor integral as a linear combination of scalar form factors and tensor structures
\begin{equation}
I^{\mu \nu}=I^2_0 g^{\mu \nu}+I^2_{11}p_1^\mu p_1^\nu+I^2_{22}p_2^\mu p_2^\nu +I^2_3\left(p_1^\mu p_2^\nu + p_2^\mu p_1^\nu\right).
\end{equation}
During this calculation, we found that we could relate all form factors of rank $n$ to form factors of rank $n-1$ or $n-2$. The most complicated matrix inversion we needed to perform in order to do this was the inversion of the matrix
\begin{equation}
  \begin{pmatrix}
    p_1^2 & p_1 \cdot p_2 \\
    p_1 \cdot p_2 & p_2^2 \\
  \end{pmatrix}\,.
\end{equation}

\begin{figure}
  \centering
  \begin{subfigure}[c]{0.32\textwidth}
    \includegraphics[width=\textwidth]{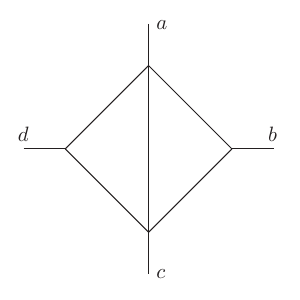}
    \subcaption{}\label{Slashed Box 1}
  \end{subfigure}
   \begin{subfigure}[c]{0.32\textwidth}
    \includegraphics[width=\textwidth]{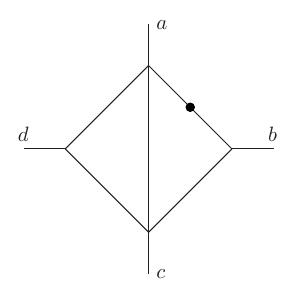}
    \subcaption{}\label{Slashed Box 2}
  \end{subfigure}
   \begin{subfigure}[c]{0.32\textwidth}
    \includegraphics[width=\textwidth]{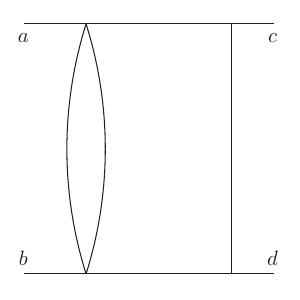}
    \subcaption{}\label{Self Energy Box}
  \end{subfigure}
  \begin{subfigure}[c]{0.32\textwidth}
    \includegraphics[width=\textwidth]{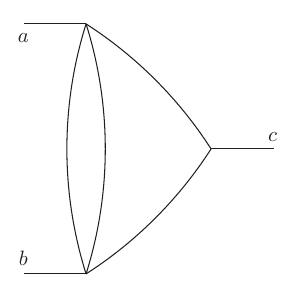}
    \subcaption{}\label{Star Trek 1}
  \end{subfigure}
  \begin{subfigure}[c]{0.32\textwidth}
    \includegraphics[width=\textwidth]{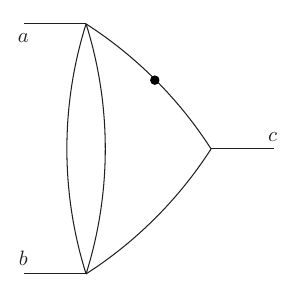}
    \subcaption{}\label{Star Trek 2}
  \end{subfigure}
    \begin{subfigure}[c]{0.32\textwidth}
    \includegraphics[width=\textwidth]{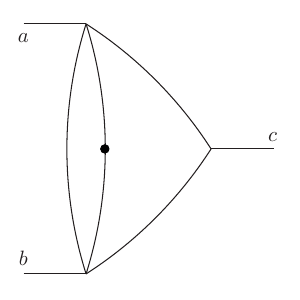}
    \subcaption{}\label{Star Trek 3}
    \end{subfigure}
   \begin{subfigure}[c]{0.32\textwidth}
    \includegraphics[width=\textwidth]{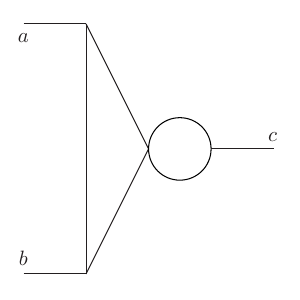}
    \subcaption{}\label{Rocket}
  \end{subfigure}
   \begin{subfigure}[c]{0.32\textwidth}
    \includegraphics[width=\textwidth]{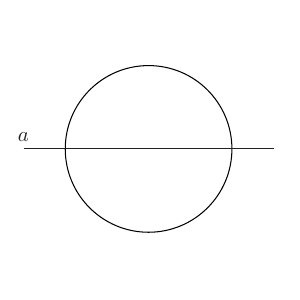}
    \subcaption{}\label{Sunset}
  \end{subfigure}
   \begin{subfigure}[c]{0.32\textwidth}
    \includegraphics[width=\textwidth]{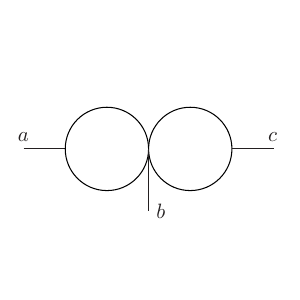}
    \subcaption{}\label{Double Loop}
  \end{subfigure}
   \begin{subfigure}[c]{0.32\textwidth}
    \includegraphics[width=\textwidth]{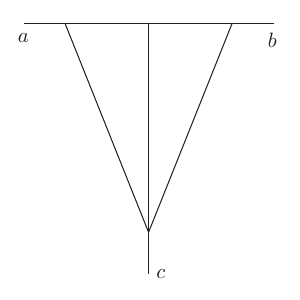}
    \subcaption{}\label{Kite}
  \end{subfigure}
   \begin{subfigure}[c]{0.32\textwidth}
    \includegraphics[width=\textwidth]{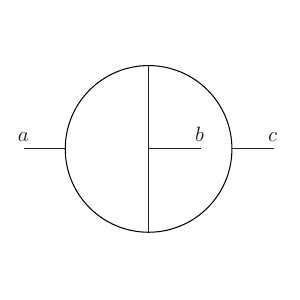}
    \subcaption{}\label{Non-Planar}
  \end{subfigure}
  \caption{The topologies for all master integrals. All external momenta are incoming. A black dot represents a power of 2 on the scalar propagator.}
  \label{fig:MasterIntegrals}
\end{figure}

After reducing the problem to the level of scalar integrals, we made use of Reduze 2 \cite{vonManteuffel:2012np} and FIRE6 \cite{Smirnov_2020} to perform an IBP reduction of the integrals. In some cases, the Feynman diagram could not be expressed in terms of a propagator basis which was conducive to IBP reduction. In these cases, once we had reduced the Feynman diagram to the level of scalar integrals, we processed these scalars using the method described in Ref.~\cite{Kvedaraite:2021ymv} such that we were left with a new set of scalar integrals in bases that were appropriate for direct application of IBP reduction. This left us with a set of Master Integrals, the topologies for which are given in Figure \ref{fig:MasterIntegrals}.
Details regarding our set of Master Integrals, as well as sources used for integral values can be found in Tables \ref{table:MasterIntegralsMOM} and \ref{table:MasterIntegrals}. We found values for most of these master integrals in  Ref.~\cite{Ussyukina_1994,Ussyukina_1995,Almeida_2010}, but for those integrals for which there are no analytical results we made use of PySecDec \cite{Borowka_2018} to evaluate them numerically.
\begin{table}
\begin{center}
\begin{tabular}{| c | c | c | c | c |}
 \hline
 Type & Source of Value & a & b & c  \\
 \hline
 Fig. \ref{Double Loop} & This Work & $2p$ & $p$ & $p$ \\
 \hline
 Fig. \ref{Star Trek 1} & \cite{Ussyukina_1994} & $-p$ & $-p$ & $2p$ \\
 \hline
 Fig. \ref{Sunset} & This Work & $2p$ & - & - \\
 \hline
 Fig. \ref{Double Loop} & This Work & $-p$ & $2p$ & $-p$ \\
 \hline
 Fig. \ref{Sunset} & This Work & $p$ & - & - \\
 \hline
 Fig. \ref{Double Loop} & This Work & $p$ & - & - \\
 \hline
 \end{tabular}
\caption{The master integrals for this work in the MOM kinematics. For master topologies, see Figure \ref{fig:MasterIntegrals}.}
\label{table:MasterIntegralsMOM}
\end{center}
\end{table}

\begin{table}
\begin{center}
\begin{tabular}{| l | l | c | c | c | c |} 
 \hline
 Topology & Source of Value  & a & b & c & d \\ 
 \hline
Fig. \ref{Slashed Box 1} & \cite{Borowka_2018} &  $-q$ & $-p_1$ & $q$ & $p_1$ \\ 
 \hline
Fig. \ref{Slashed Box 1} & \cite{Borowka_2018} &  $-q$ & $-p_1$ & $q$ & $p_1$ \\ 
 \hline
Fig. \ref{Slashed Box 1} & \cite{Borowka_2018} &  $-p_1$ & $p_2$ & $2p_1-p_2$ & $-p_1$ \\ 
 \hline
Fig. \ref{Slashed Box 2} & \cite{Borowka_2018}&  $q$ & $-p_1$ & $-q$ & $p_1$ \\ 
 \hline
Fig. \ref{Slashed Box 2} & \cite{Borowka_2018} &  $-p_1$ & $p_2$ & $2p_1-p_2$ & $-p_1$ \\ 
 \hline
Fig. \ref{Slashed Box 2} & \cite{Borowka_2018} &  $2p_1-p_2$ & $-p_1$ & $-p_1$ & $p_2$ \\ 
 \hline
Fig. \ref{Slashed Box 2} & \cite{Borowka_2018} &  $-p_1$ & $-p_1$ & $2p_1-p_2$ & $p_2$ \\ 
 \hline
Fig. \ref{Slashed Box 2} & \cite{Borowka_2018} &  $2p_1-p_2$ & $p_2$ & $-p_1$ & $-p_1$ \\ 
 \hline
Fig. \ref{Self Energy Box} & \cite{Borowka_2018} &  $-q$ & $p_1$ & $-p_1$ & $-q$ \\ 
 \hline
Fig. \ref{Self Energy Box} & \cite{Borowka_2018} &  $-p_1$ & $2p_1-p_2$ & $-p_1$ & $p_2$ \\ 
 \hline
Fig. \ref{Star Trek 1} & \cite{Ussyukina_1994} and \cite{Borowka_2018}  &  $-q$ & $2p_1-p2$ & $-p_1$ & - \\ 
 \hline
Fig. \ref{Star Trek 1} & \cite{Ussyukina_1994} and \cite{Borowka_2018} &  $p_1$ & $-p_2$ & $-q$ & - \\ 
 \hline
Fig. \ref{Star Trek 1} & \cite{Ussyukina_1994} and \cite{Borowka_2018}  &  $2p_1$ & $p_2-2p_1$ & $-p_2$ & - \\ 
 \hline
Fig. \ref{Star Trek 1} & \cite{Ussyukina_1994} and \cite{Borowka_2018} &  $2p_1$ & $-p_1$ & $-p_1$ & - \\ 
 \hline
Fig. \ref{Star Trek 1} & \cite{Ussyukina_1994} and \cite{Borowka_2018}  &  $-p_1$ & $-p_1$ & $2p_1$ & - \\ 
 \hline
Fig. \ref{Star Trek 1} & \cite{Ussyukina_1994} and \cite{Borowka_2018}  &  $-p_2$ & $p_2-2p_1$ & $2p_1$ & - \\ 
 \hline
Fig. \ref{Star Trek 2} & \cite{Ussyukina_1994} and \cite{Borowka_2018} &  $p_1$ & $-p_2$ & $-q$ & - \\ 
 \hline
Fig. \ref{Star Trek 2} & \cite{Ussyukina_1994} and \cite{Borowka_2018} &  $-q$ & $2p_1-p_2$ & $-p_1$ & - \\ 
 \hline
Fig. \ref{Star Trek 2} & \cite{Ussyukina_1994} and \cite{Borowka_2018} &  $p_2-2p_1$ & $p_1$ & $q$ & - \\ 
 \hline
Fig. \ref{Star Trek 2} & \cite{Ussyukina_1994} and \cite{Borowka_2018} &  $-p_2$ & $p_2-2p_1$ & $2p_1$ & - \\ 
 \hline
Fig. \ref{Star Trek 2} & \cite{Ussyukina_1994} and \cite{Borowka_2018} &  $-2p_1$ & $2p_1-p_2$ & $p_2$ & - \\ 
 \hline
Fig. \ref{Star Trek 3} &  \cite{Ussyukina_1994} and \cite{Borowka_2018} &  $-q$ & $2p_1-p_2$ & $-p_1$ & - \\ 
 \hline
Fig. \ref{Star Trek 3} & \cite{Ussyukina_1994} and \cite{Borowka_2018} &  $p_2$ & $q$ & $-p_1$ & - \\ 
 \hline
Fig. \ref{Star Trek 3} & \cite{Ussyukina_1994} &  $2p_1$ & $p_2-2p_1$ & $-p_2$ & - \\ 
 \hline
Fig. \ref{Rocket} & This Work, \cite{Ussyukina_1995} and \cite{Borowka_2018} &  $p_1$ & $q$ & $-p_2$ & - \\ 
   \hline
Fig. \ref{Sunset} & This Work &  $p_1$ & - & - & - \\ 
 \hline
Fig. \ref{Sunset} &  This Work &  $2p_1$ & - & - & - \\ 
 \hline
Fig. \ref{Sunset} & This Work &  $2p_1-p_2$ & - & - & - \\ 
 \hline
Fig. \ref{Double Loop} & This Work &  $p_1$ & 0 & $-p_1$ & - \\ 
 \hline
Fig. \ref{Double Loop} & This Work  & $2p_1$ & $-p_1$ & $-p_2$ & - \\ 
 \hline
Fig. \ref{Kite} & \cite{Ussyukina_1994} &  $q$ & $-p_1$ & $p_2$ & -\\ 
 \hline
Fig. \ref{Kite} & \cite{Ussyukina_1994} &  $2p_1$ & $-p_2$ & $p_2-2p_1$ & - \\ 
 \hline
Fig. \ref{Kite} & \cite{Ussyukina_1994} &  $-q$ & $-p_1$ & $2p_1-p_2$ & - \\ 
 \hline
Fig. \ref{Non-Planar} & \cite{Ussyukina_1994} &  $p_1$ & $-q$ & $-p_2$ & - \\ 
 \hline
\end{tabular}
\caption{The master integrals for this work in the SMOM kinematics. For master topologies, see Figure \ref{fig:MasterIntegrals}.}
\label{table:MasterIntegrals}
\end{center}
\end{table}

We found a typo in eq. (29) of \cite{Ussyukina_1994} where an extra factor of $\frac{2\pi^2}{3}$ is present in the last line and hence we used the integral definition in eq. (28). Moreover, eq. (11) breaks down in some kinematic configuration, e.g.\ $x=4$ and $y=1$, and as a result eq. (9) needs to be evaluated numerically.

In some cases, we needed to improve the analytical expression of the integral found in the literature and in order to do so we made use of PySecDec to evaluate the missing parts in the $\epsilon$-expansion numerically.

We have been able to retrieve the scale dependence of these integrals (which is logarithmic) by exploiting the fact that there is only one independent kinematic scale. This meant that we could rescale the master integral, leaving only dimensionless loop and external momenta before either performing the integration, or using PySecDec to perform the calculation. To illustrate, consider the loop integral $I_{ex}$, with dimensionfull quantities $k$, a loop integral, and $p$, an external momentum. If we re-express these in terms of a dimensionfull scale, $\nu$, and dimensionless variables, $\tilde{k}$ and $\tilde{p}$,
\begin{align}
  k &= \nu \tilde{k} \\
  p &= \nu \tilde{p}
\end{align}
we retrieve
\begin{align}
  I_{ex} &= (\mu^2)^\epsilon \int \frac{d^dk}{(2\pi)^d}\frac{1}{[k^2]^a[(k-p)^2]b} \\
  &= \frac{1}{(\nu^2)^{a+b-2}}\frac{(\mu^2)^\epsilon}{(\nu^2)^\epsilon} \int \frac{d^d\tilde{k}}{(2\pi)^d}\frac{1}{[\tilde{k}^2]^a[(\tilde{k}-\tilde{p})^2]b}.
\end{align}
As can be seen, we retain the overall mass dimension of the integral in the first prefactor, but are then able to retrieve the log behaviour of the integral through the second prefactor, upon expanding in $\epsilon$. 
\newpage
\cleardoublepage
\section{Results and numerics}
\label{sec:results--numerics}
In this section we present our results for the conversion factors between the $\overline{\mathrm{MS}}$ and the $\mom$, $\momb$ and $\smomb$ schemes.
The respective conversion factors exhibit a scale dependence that mirrors the scale dependence of the relevant semi-leptonic Wilson coefficient.
Working in renormalization group improved perturbation theory, this scale dependence will cancel order-by-order in perturbation theory for the product of the Wilson coefficient and a given conversion factor.
By studying the residual scale dependence of this product, we can estimate the uncertainty from unknown higher-order corrections.

\subsection{Scale dependence}
In the following we will combine the matching corrections with the renormalization-group-improved Wilson coefficient.
The results of our two-loop calculation together with \eqref{eq:Cfmaster} and \eqref{eq:COmaster} determine the conversion factor $\mathcal{C}_O^{\overline{\mathrm{MS}}\to i}$ including its analytic dependence on the
$\overline{\rm MS}$ scale $\mu$.
The fact that the product of the Wilson coefficient $C^{\overline{\mathrm{MS}}}(\mu)$ of the semi-leptonic operator \eqref{Effective-semileptonic-operator} and the conversion factor is scale-independent implies the renormalization group equation
\begin{equation}
\label{scale independence of the product}
\frac{d}{d\; {\rm ln}(\mu)}\mathcal{C}_O^{\overline{\mathrm{MS}}\to i}(\mu)=-\gamma_W \mathcal{C}_O^{\overline{\mathrm{MS}}\to i},
\end{equation}
where 
$\gamma_W$ is the anomalous dimension of the semi-leptonic operator.
The two-loop anomalous dimension and the one-loop electroweak matching corrections for the Wilson at the electroweak scale are given in Ref.~\cite{Brod:2008ss}.
This renormalization-group equation provides an additional check of our calculation.
In the following, we study the residual scale dependence in the product $C_O^{\overline{\mathrm{MS}}}(\mu_L) \cdot \mathcal{C}_O^{\overline{\mathrm{MS}}\to i}(\mu_L)$, where $\mu_L \sim O({\rm GeV})$ is the lattice scale and the large logarithms $\ln \mu_L/M_W$ are summed in renormalization-group-improved perturbation theory.

To this end we write the Wilson coefficient at the lattice scale as a product
\begin{equation}
\label{wilson coefficient evolution}
C_O(\mu_L)=\mathcal{U}(\mu_L,\mu_W)\;C_O(\mu_W)
\end{equation}
of the Wilson coefficient at the electroweak scale and the evolution kernel $\mathcal{U}(\mu,\mu_0)$, where $\mu_W$ is the weak scale, and where the evolution kernel fulfils the renormalization group equation
\begin{equation}
\label{rge evolution kernel}
\frac{d}{d\; {\rm ln}(\mu)}\mathcal{U}(\mu,\mu_0)=\gamma_W\;\mathcal{U}(\mu,\mu_0).
\end{equation}
In the following we will consider the expanded anomalous dimension
\begin{equation}
\label{anomalous dimension}
\gamma_W=\gamma_W^{(0)}\frac{\alpha}{4\pi}+\gamma_W^{(1)}\frac{\alpha\alpha_s}{(4\pi)^2} + \gamma_W^{(2)}\frac{\alpha\alpha_s^2}{(4\pi)^3},
\end{equation}
up to $\mathcal{O}(\alpha \alpha_s^2)$, since the value of $\gamma_W^{(2)}$ is sensitive to $O(\alpha \alpha_s)$ scheme transformation of the effective theory. The value $ \gamma_W^{(2)}$ is currently not known, but will play a part in our numerical analysis later.

The solution of \eqref{rge evolution kernel}, under the assumption that $\alpha(\mu)=\alpha(\mu_W)=\alpha$ is scale independent, is found to be
\begin{equation}
\label{analytical expansion U}
\begin{split}
\mathcal{U}(\mu,\mu_0)=&1+\frac{\alpha}{4\pi}\left(\gamma^{(0)}_W\;{\rm ln}\left(\frac{\mu}{\mu_0}\right)\right)-\frac{\alpha}{4\pi}\frac{1}{2\beta_0^{(N_f)}}\gamma_W^{(1)}\left({\rm ln}\left(\frac{\alpha_s(\mu)}{\alpha_s(\mu_0)}\right)\right)\\
&+\frac{\alpha}{4\pi}\frac{1}{2\beta_0^{(N_f)}}\left(\gamma_W^{(1)}\frac{\beta_1^{(N_f)}}{\beta_0^{(N_f)}}-\gamma_W^{(2)}\right)\frac{\alpha_s(\mu)-\alpha_s(\mu_0)}{4\pi},
\end{split}
\end{equation}
where $\beta^{(N_f)}_0=11\;C_A/3-(4/3)T_F\;N_f$ and $\beta^{(N_f)}_1=(34/3)\;C_A^2-4\;C_F\;T_F\;N_f-(20/3)\;C_A\;T_F\;N_f$ are the first two terms of the QCD beta function that determines the running of $\alpha_s(\mu)$; $C_F=\frac{4}{3}$ is the Casimir factor of the fundamental representation of $SU(3)$, $C_A=3$ is the Casimir factor of the adjoint representation of $SU(3)$ and $T_F=\frac{1}{2}$ is the trace normalization of the fundamental representation, while $N_f$ is the number of flavours.

When evolving the Wilson coefficient down to the lattice scale we must take into account the threshold corrections due to the fact that we integrate out the heavy quarks. In order to do this, we split the evolution in three steps:
\begin{itemize}
\item $\mu_W \rightarrow\mu_b\sim 4.4 \; {\rm GeV}$, here we have $N_f=5$;
\item $\mu_b \rightarrow\mu_L\sim 1-6 \; {\rm GeV}$, here we have $N_f=4$.
\end{itemize} 

Combining these steps and expanding in powers of $\alpha$ and $\alpha_s$ as
\begin{equation}
\begin{split}
&C^{\overline{\mathrm{MS}}}_O(\mu_W)=1+\frac{\alpha}{4\pi}\;C_O^e\left(\mu_W,M_Z\right)+\frac{\alpha}{4\pi}\frac{\alpha_s}{4\pi}\;C_O^{es}\left(\mu_W,M_Z\right),\\
&\mathcal{C}_O^{\overline{\mathrm{MS}}\to i}(\mu_L)=1+\frac{\alpha_s(\mu_L)}{4\pi}\;\mathcal{C}_O^{s,1}\left(-p^2,\mu_L^2\right)+\frac{\alpha_s^2(\mu_L)}{(4\pi)^2}\;\mathcal{C}_O^{s,2}\left(-p^2,\mu_L^2\right)\\
&+\frac{\alpha}{4\pi}\;\mathcal{C}_O^{e}\left(-p^2,\mu_L^2\right)+\frac{\alpha}{4\pi}\frac{\alpha_s(\mu_L)}{4\pi}\;\mathcal{C}_O^{es}\left(-p^2,\mu_L^2\right),
\end{split}
\end{equation}
we obtain the final expression of the low-scale matching,
\begin{equation}
\label{matching high low scale}
C^i_O(\mu_L,p^2)=C^{\overline{\mathrm{MS}}}_O(\mu_W)\;\mathcal{U}(\mu_W,\mu_L)\;\mathcal{C}_O^{\overline{\mathrm{MS}}\to i}(\mu_L,p^2)=C^i_\alpha + C^i_{\alpha_s}+\frac{\alpha}{4\pi}\left(C^i_{\alpha,\alpha_s\; \rm LL}+C^i_{\alpha,\alpha_s\; \rm NLL}\right),
\end{equation}
where the superscript $i$ denotes either of the RI schemes, $C^i_\alpha$ is the resummed QED contribution, 
$C^i_{\alpha_s}$ is the leading $O(\alpha_s)$ effect and $C^i_{\alpha,\alpha_s\; \rm LL}$ and $C^i_{\alpha,\alpha_s\; \rm NLL}$ are the leading-log (LL) and next-to-leading-log (NLL) strong corrections to the electromagnetic contributions,
\begin{equation}
\label{explicit matching expressions}
\begin{split}
&C^i_\alpha=1+\frac{\alpha}{4\pi}\left(C_O^e(\mu_W,M_Z)+\gamma^{(0)}_W\;{\rm ln}\left(\frac{\mu_L}{\mu_W}\right)+\mathcal{C}_O^e(-p^2,\mu_L^2)\right),\\
&C^i_{\alpha_s}=\frac{\alpha_s(\mu_L)}{4\pi}\left(\mathcal{C}_O^{s,1}\left(-p^2,\mu_L^2\right)+\frac{\alpha_s(\mu_L)}{4\pi}\;\mathcal{C}_O^{s,2}\left(-p^2,\mu_L^2\right)\right),\\
&C^i_{\alpha,\alpha_s\; \rm LL}=-\gamma^{(1)}_W\left(\frac{1}{2\beta_{(0)}^{(4)}}{\rm ln}\left(\frac{\alpha_s(\mu_L)}{\alpha_s(\mu_b)}\right)+\frac{1}{2\beta_{(0)}^{(5)}}{\rm ln}\left(\frac{\alpha_s(\mu_b)}{\alpha_s(\mu_W)}\right)\right),\\
&C^i_{\alpha,\alpha_s\; \rm NLL}=\frac{\alpha_s(\mu_L)}{4\pi}\left(\mathcal{C}_O^{es}\left(-p^2,\mu_L^2\right)+\bar{\gamma}^{(4)}\right)\\
&+\frac{\alpha_s(\mu_b)}{4\pi}\left(\bar{\gamma}^{(5)}-\bar{\gamma}^{(4)}\right)+\frac{\alpha_s(\mu_W)}{4\pi}\left(C_O^{es}(\mu_W,M_Z)-\bar{\gamma}^{(5)}\right),
\end{split}
\end{equation}
with $\bar{\gamma}^{(N_f)}=\frac{1}{2\beta_0^{(N_f)}}\left(\gamma_W^{(1)}\;\frac{\beta_1^{(N_f)}}{\beta_0^{(N_f)}}-\gamma_W^{(2)}\right)$.
We stress again that the contributions coming from $C^i_{\alpha_s}$ will vanish to all orders for the renormalization schemes defined by the conditions \eqref{eq:SMOMprojcond} and \eqref{eq:MOMprojcond} by virtue of the Ward Identity. 

\subsection{$\mom$ \& $\momb$}
\label{subs: RI-MOM}
Here and in the following, we will collect the results for the different operator and field conversion factors.
The explicit form of the operator conversion factor will depend on the renormalization kinematics and on the choice of projectors.
Contrary to this, the field conversion factor is the same in the $\mom$, $\smom$, $\momb$, and $\smomb$ schemes and reads
\begin{equation}
\label{field conversion factor}
\begin{split}
&\mathcal{C}_f^{{\rm RI} \to \overline{\mathrm{MS}}}=1+\frac{\alpha_s}{4\pi}\;C_F \; \xi_{s}\left({\rm ln} \left(\frac{-p^2}{\mu^2}\right)-1\right)+\frac{\alpha}{4\pi}\; \xi \;q^2 \left({\rm ln} \left(\frac{-p^2}{\mu^2}\right)-1\right)\\
&+\frac{\alpha}{4\pi}\frac{\alpha_s}{4\pi}\frac{1}{4}C_F\;q^2 \left(4 \;{\rm ln}\left(\frac{-p^2}{\mu^2}\right) \left(\xi\;\xi_{s}\; {\rm ln}\left(\frac{-p^2}{\mu^2}\right)-2\;\xi\;\xi_{s}-3\right)+8\;\xi\;\xi_{s}+5\right)\\
&+\left(\frac{\alpha_s}{4\pi}\right)^2 \frac{1}{8} \; C_F \;\Bigg(C_A \left(-9 \;\xi_{s}^2-52 \;\xi_{s}+24 \left(\xi_{s}+1\right) \zeta (3)-82\right)+C_F \left(8 \;\xi_{s}^2+5\right)+28 \;N_f \;T_F\\
&+2\;{\rm ln}\left(\frac{-p^2}{\mu^2}\right) \left(2\;C_A\;\xi_{s} \left(\xi_{s}+3\right) {\rm ln}\left(\frac{-p^2}{\mu^2}\right)+C_A \left(\xi_{s} \left(\xi_{s}+8\right)+25\right)-6\;C_F-8\;N_f\;T_F\right)\Bigg)\\
&+ O\left(\left( \frac{\alpha_s}{4\pi}\right)^3\right),
\end{split}
\end{equation}
where $\xi\;(\xi_s)$ is the photon (gluon) gauge parameter.
The two-loop QCD correction is given in Ref.~\cite{Gracey_2003} while the $\mathcal{O}(\alpha \alpha_s)$ contribution is a novel result.

For MOM kinematics we give the operator conversion factor both for the $\mom$ scheme (using the projector in Eq. \eqref{eq:projstd}) and our new $\momb$ scheme using the projector in Eq. \eqref{RI'-MOM projector}.
Here we used the results in Ref.~\cite{Gracey_2003} to implement the two-loop QCD corrections to the operator conversion factors.
For the $\mom$ scheme we find
\begin{equation}
\label{conversion MOM old projector}
\begin{split} 
&\mathcal{C}_O^{\overline{\mathrm{MS}}\to \mom}=1+\frac{\alpha}{4\pi}\left(-2{\rm ln}\left(\frac{-p^2}{\mu^2}\right)-1.23728\;\xi-0.87851 \right)+\frac{\alpha_s(\mu_L)}{4\pi}\left(-0.5\;  C_F\;\xi_{s}\right)\\
&+\frac{\alpha}{4\pi}\frac{\alpha_s(\mu_L)}{4\pi}\;C_F\left(\left(\xi_{s}+\frac{3}{2}\right) {\rm ln}\left(\frac{-p^2}{\mu^2}\right)+\xi \left(-0.13990\; \xi_{s}-1.38558 \right)+0.56651   \xi_{s}+6.96381\right)\\
&+\left(\frac{\alpha_s(\mu_L)}{4\pi}\right)^2 \frac{1}{8}\;C_F \left(-C_A \left(\xi_{s}\left(3\; \xi_{s}+14\right)+25\right)+6\;C_F +8\;N_f\;T_F\right) + O\left(\left( \frac{\alpha_s(\mu_L)}{4\pi}\right)^3\right).
\end{split}
\end{equation}
As a check, we combine this result with the scheme conversion factor (\ref{converion MS W-Reg}) and find agreement with the one-loop QED result obtained in Ref.~\cite{DiCarlo:2019thl} (in our convention $\xi=1(0)$ in the Feynman (Landau) gauge).
\begin{figure}
\centering
\includegraphics[scale=1]{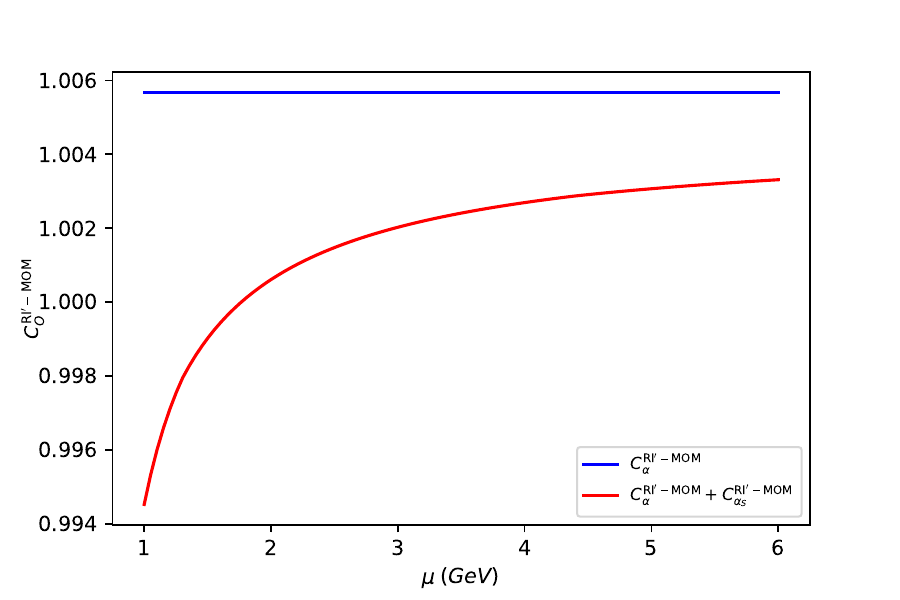}
\caption{Residual $\mu$-dependence of the low-scale Wilson coefficient $C_O^{\mom}$
for the conventional projector
choice, where we set $\xi=\xi_s=0$ and $-p^2=9$. It is evident that the leading strong corrections introduce an artificial scale dependence at low scales $\mu\sim \mu_L$. }
\label{fig: running matching RIMOM old}
\end{figure}

The choice of the conventional projector leads to the presence of $O(\alpha_s)$ and $O(\alpha_s^2)$ corrections, 
where the latter correction does not even vanish for $\xi=0$.
The $\mu$ dependence of $\alpha_s$ will introduce an artificial scale dependence into this conversion factor that is not related to the operator anomalous dimension.
The resulting residual dependence of $C_O^{\mom}(\mu_L,p^2)$ on the scale $\mu_L$, shown of Figure~\ref{fig: running matching RIMOM old}, suggest at least an uncertainty of $\pm 0.5\%$ from unknown three-loop QCD corrections.

For the $\momb$ scheme using the projector which we derived (Eq. \eqref{RI'-MOM projector}), we find

\begin{equation}
\label{conversion MOM new projector}
\begin{split}
&\mathcal{C}_O^{\overline{\mathrm{MS}}\to \momb}=1+\frac{\alpha}{4\pi}\left(-2\;{\rm ln} \left(\frac{-p^2}{\mu^2}\right)-1.51506\; \xi-0.87851\right)\\
&+\frac{\alpha}{4\pi}\frac{\alpha_s(\mu_L)}{4\pi}\;C_F\left(\frac{3}{2}\;{\rm ln}\left(\frac{-p^2}{\mu^2}\right)+\xi(-0.52364\; \xi_{s}-1.23469)+0.31714 \; \xi_{s}+6.04812\right)\\
&+ O\left(\frac{\alpha}{4\pi} \left( \frac{\alpha_s(\mu_L)}{4\pi}\right)^2\right),\\
\end{split}
\end{equation}
where we have explicitly checked that all pure QCD corrections vanish.
 
\begin{figure}
\centering
\includegraphics[scale=1]{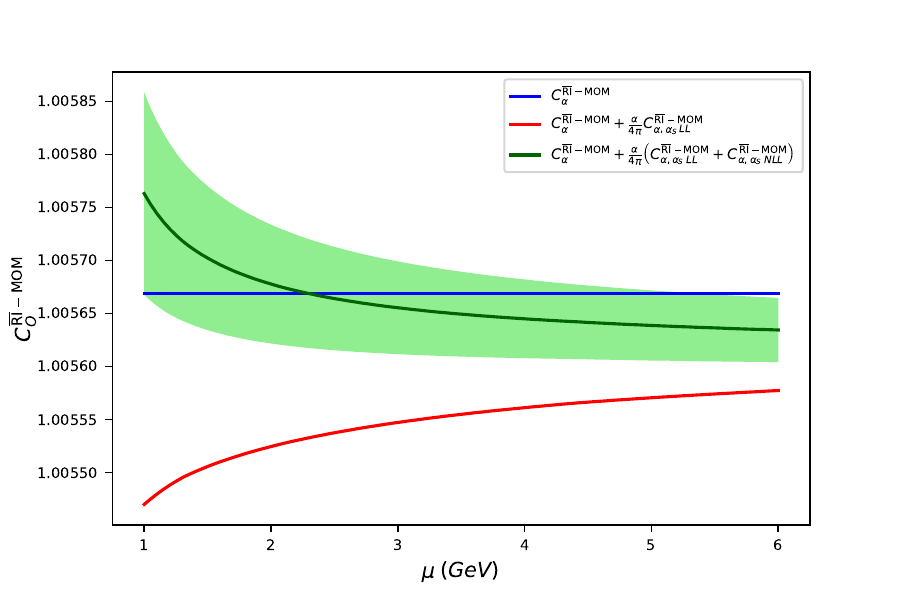}
\caption{Residual scale dependence of the low-scale Wilson coefficient $C_O^{\momb}$,
where we set $\xi=\xi_s=0$, $-p^2=9$ and $\gamma_W^{(2)}=0$ (dark green curve). We can see the reduction of
the scale dependence in going from $C^{\momb}_{\alpha,\alpha_s LL}$ to $C^{\momb}_{\alpha,\alpha_s NLL}$, while
the residual scale dependence is dramatically reduced compared to the $\mom$ scheme (Figure \ref{fig: running matching RIMOM old}). The light green shaded area shows the effect of the unknown value of $\gamma_W^{(2)}$ on the
next-to-leading-log contribution: the upper limit is obtained with $\gamma_W^{(2)}=-100$, while the lower limit is given by $\gamma_W^{(2)}$=100.}
\label{fig: running matching RIMOM new}
\end{figure}

Combining the expression \eqref{matching high low scale} with the values of the coefficients given by \eqref{conversion MOM new projector} and \cite{Brod:2008ss} we obtain the expression of the Wilson Coefficient in the $\momb$ scheme
\begin{equation}
\label{Wilson Coefficient MOM}
\begin{split}
&C_O^{\momb}(\mu_L,p^2)=1+\frac{\alpha}{4\pi}\left(-\frac{11}{3}-1.51506\;\xi-0.87851-2\;{\rm ln}\left(\frac{-p^2}{M_Z^2}\right)\right)
\\
&+\frac{\alpha}{4\pi}\left(-4\left(\frac{1}{2\beta_{(0)}^{(4)}}\;{\rm ln}\left(\frac{\alpha_s(\mu_L)}{\alpha_s(\mu_b)}\right)+\frac{1}{2\beta_{(0)}^{(5)}}\;{\rm ln}\left(\frac{\alpha_s(\mu_b)}{\alpha_s(\mu_W)}\right)-\frac{\alpha_s(\mu_L)}{8\pi}\;{\rm ln}\left(\frac{-p^2}{\mu_L^2}\right) \right)\right)\\
&+\frac{\alpha}{4\pi}\left(\frac{\alpha_s(\mu_L)}{4\pi}\left(C_F\left(\;(-0.52364\; \xi_{s}-1.23469)\;\xi+0.31714 \; \xi_{s}+6.04812\right)+\bar{\gamma}^{(4)}\right)\right.\\
&\left.+\frac{\alpha_s(\mu_b)}{4\pi}\left(\bar{\gamma}^{(5)}-\bar{\gamma}^{(4)}\right)+\frac{\alpha_s(\mu_W)}{4\pi}\left(C_O^{es}(\mu_W,M_Z)-\bar{\gamma}^{(5)}\right)\right).
\end{split}
\end{equation}
We recall that $\bar{\gamma}^{(N_f)}=\frac{1}{2\beta_0^{(N_f)}}\left(\gamma_W^{(1)}\;\frac{\beta_1^{(N_f)}}{\beta_0^{(N_f)}}-\gamma_W^{(2)}\right)$.

We can see that there is an exact cancellation of the $\mu$ dependence in the case of pure electromagnetic corrections.

When we switch on QCD interactions, however, while the scale dependence of the LL term given by ${\rm ln}\left(\frac{\alpha_s(\mu_L)}{\alpha_s(\mu_b)}\right)$ cancels with the explicit scale dependence of the NLL terms in the conversion factor, there is a residual scale dependence coming from the other NLL terms proportional to $\frac{\alpha_s(\mu_L)}{4\pi}\left(\bar{\gamma}^{(4)}+\mathcal{C}_{O}^{es}\right)$. This effect, nevertheless, is sub-leading with respect to the residual scale dependence in \eqref{conversion MOM old projector}, which is dominant at the low scale. In order to explore the contributions from higher order corrections in our numerical analysis, we vary $ \gamma_W^{(2)}$ in the range $-100 < \gamma_W^{(2)} < 100$, since it is currently not known. This range was chosen to reflect the typical range of values for this type of quantity. This effect can be seen in the light green bands of Figs. \ref{fig: running matching RIMOM new} and \ref{fig: running matching RISMOM}

The resulting residual scale dependence of $C^{\momb}_O(\mu_L,p^2)$ is given in Figure \ref{fig: running matching RIMOM new} and suggest a very small, i.e.\ $\pm 2 \cdot 10^{-4}$, uncertainty from higher order corrections. While these uncertainty are encouraging, we would like to remind the reader that the uncertainties from higher order electroweak corrections are not included and that the contribution of $\gamma_W^{(2)}$ to the NLL QCD corrections is only estimated.

\subsection{$\smomb$}
\label{subs: RI-SMOM}

The field conversion factor for $\smomb$ equals the one for the other RI schemes given in eq.~\eqref{field conversion factor}.
For the operator conversion factor we only consider the $\smomb$ scheme (i.e.\ the one defined using the projector seen in \eqref{RI-SMOM projector}), which should not yield any pure QCD corrections.
It reads
\begin{equation}
\label{conversion SMOM}
\begin{split}
&\mathcal{C}_O^{\overline{\mathrm{MS}}\to \smomb}=1+\frac{\alpha}{4\pi}\left(-2\;{\rm ln} \left(\frac{-p^2}{\mu^2}\right)-1.62969 \;\xi-1.54518\right)\\
&+\frac{\alpha}{4\pi}\frac{\alpha_s}{4\pi}\;C_F\left(\frac{3}{2}\;{\rm ln}\left(\frac{-p^2}{\mu^2}\right)+\xi(-0.18563\; \xi_{s}-0.24468)+0.58741\; \xi_{s}+5.69043\right)\\
&+ O\left(\frac{\alpha}{4\pi} \left( \frac{\alpha_s}{4\pi}\right)^2\right) \,,
\end{split}
\end{equation}
where we have again checked explicitly that $O\left( \left( \frac{\alpha_s}{4\pi}\right)^2\right)$ corrections are absent. If we use the $x$-dependent projector, our result then becomes
\begin{multline}
    \mathcal{C}_O^{\overline{\mathrm{MS}}\to \smomb}=1+\frac{\alpha}{4\pi}\left(-2\;{\rm ln} \left(\frac{-p^2}{\mu^2}\right)-1.62969 \;\xi-0.520868 x-1.28474\right)\\
    +\frac{\alpha}{4\pi}\frac{\alpha_s}{4\pi}\;C_F\bigg(\frac{3}{2}\;{\rm ln}\left(\frac{-p^2}{\mu^2}\right)+\xi((0.376692 x-0.373979)\; \xi_{s}+0.00885096x-0.249106)+\\
    +(0.462791x+0.356019)\; \xi_{s}-1.22351x+6.30218\bigg)+ O(\frac{\alpha}{4\pi} \left( \frac{\alpha_s}{4\pi}\right)^2) \,,
\end{multline}
The two-loop QCD computation needed for this check can be extracted from intermediate results that appear in the QCD renormalization of the vector current in the SMOM scheme \cite{Gracey_2011}.
\begin{figure}
\centering
\includegraphics[scale=1]{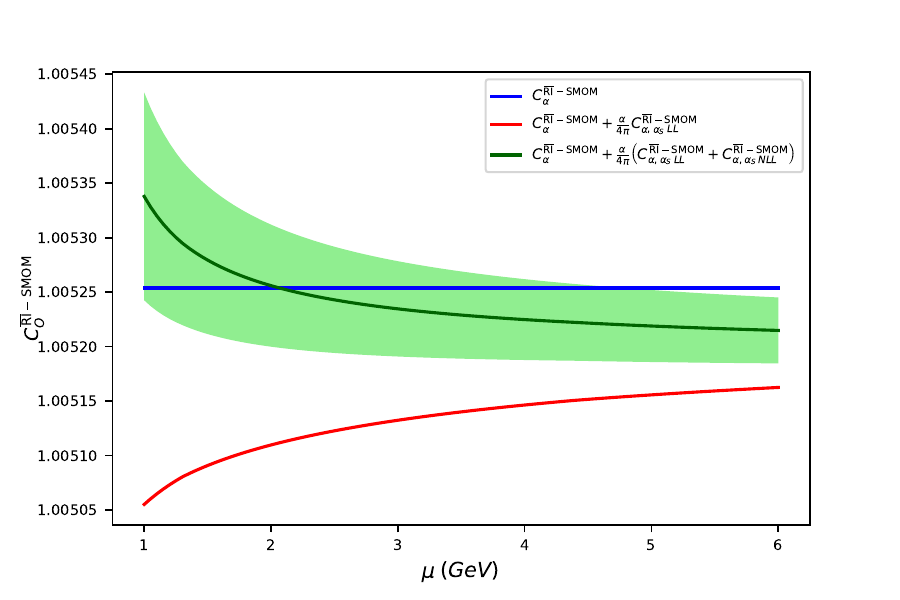}
\caption{Residual scale dependence of the low-scale Wilson coefficient $C_O^{\smomb}$,
where we set $\xi=\xi_s=0$, $-p^2=9$ and $\gamma_W^{(2)}=0$ (dark green curve). As in Figure \ref{fig: running matching RIMOM new}, the residual scale dependence is suppressed by $\alpha$. The light green shaded area shows the effect of the value of $\gamma_W^{(2)}$ on the NLL contribution: the top limit is obtained with $\gamma_W^{(2)}=-100$, while the bottom limit is given by $\gamma_W^{(2)}$=100.}
\label{fig: running matching RISMOM}
\end{figure}
\newpage
As in \ref{subs: RI-MOM}, we can combine the analytical expression \eqref{wilson coefficient evolution} with the numerical values in \eqref{conversion SMOM} and \cite{Brod:2008ss} obtaining the Wilson Coefficient in the SMOM scheme

\begin{equation}
\label{Wilson Coefficient SMOM}
\begin{split}
&C_O^{\smomb}(\mu_L,p^2)=1+\frac{\alpha}{4\pi}\left(-\frac{11}{3}-1.62969 \;\xi-1.54518-2\;{\rm ln}\left(\frac{-p^2}{M_Z^2}\right)\right)
\\
&+\frac{\alpha}{4\pi}\left(-4\left(\frac{1}{2\beta_{(0)}^{(4)}}\;{\rm ln}\left(\frac{\alpha_s(\mu_L)}{\alpha_s(\mu_b)}\right)+\frac{1}{2\beta_{(0)}^{(5)}}\;{\rm ln}\left(\frac{\alpha_s(\mu_b)}{\alpha_s(\mu_W)}\right)-\frac{\alpha_s(\mu_L)}{8\pi}\;{\rm ln}\left(\frac{-p^2}{\mu_L^2}\right) \right)\right)\\
&+\frac{\alpha}{4\pi}\left(\frac{\alpha_s(\mu_L)}{4\pi}\left(\;C_F\left(\;(-0.18563\; \xi_{s}-0.24468)\;\xi+0.58741\; \xi_{s}+5.69043\right)+\bar{\gamma}^{(4)}\right)\right.\\
&\left.+\frac{\alpha_s(\mu_b)}{4\pi}\left(\bar{\gamma}^{(5)}-\bar{\gamma}^{(4)}\right)+\frac{\alpha_s(\mu_W)}{4\pi}\left(C_O^{es}(\mu_W,M_Z)-\bar{\gamma}^{(5)}\right)\right).
\end{split}
\end{equation}

The resulting residual scale dependence of $C_O^{\smomb}(\mu_L,p^2)$ is given in Figure \ref{fig: running matching RISMOM} and again suggests a very small, i.e.\ $\pm 2 \cdot 10^{-4}$, uncertainty from higher order corrections.
Again, the uncertainties from higher order electroweak corrections are not included, while the contribution of $\gamma_W^{(2)}$ to the NLL QCD corrections is only estimated.

\section{Conclusions}
\label{sec:conclusions}
In this paper we have calculated the scheme conversions for the semi-leptonic weak effective operator between the $\overline{\rm{MS}}$ scheme and the $\mom$ scheme, as well as to our newly defined $\momb$ and $\smomb$ schemes.
We emphasized the importance of the projector in the definition of these schemes and found that a conventional choice of projector leads to an artificial QCD correction to the conversion factor with a bad perturbative convergence.
Using the Ward identity in the pure-QCD limit, we defined modified vesions of
 the $\mom$ and $\smom$ schemes that rectify this problem, by showing the existence of adequate new projectors, thereby defining the $\momb$ and $\smomb$ schemes.
Performing an effective field theory analysis with renormalization-group-improved perturbation theory we showed that these schemes indeed exhibit an excellent perturbative convergence,
when LL and partial NLL QCD corrections were added to the photonic corrections. We exhibited the dependence on
the various thresholds $\mu_W$, $\mu_b$ and the scheme conversion (lattice matching) scale
$\mu_L$ and studied the cancellation of the $\mu_L$ dependence for the product of the $\overline{\rm MS}$
Wilson coefficient and the conversion factor (or equivalently, for the RI Wilson coefficients or RI operator matrix elements).

Given the theoretical attractiveness and good perturbative convergence, we argue that the schemes defined with our proposed projectors should be used in future work on semileptonic decays in place of the coventional RI schemes.
In particular, this should allow a better precision in determining CKM matrix elements in future phenomenological
analyses.

Our approach also lends itself to a systematic improvement of short-distance contributions. We leave
the evaluation of the relevant three-loop anomalous dimensions and two-loop electroweak matching calculations required
for completing the NLL QCD corrections for future work.

\section*{Acknowledgments}
The authors would like to thank Sandra Kvedarait{\.e} for helpful discussions as well as advice on the use of various
computer programs, and John Gracey, Paul Rakow and Francesco Sanfilippo for useful discussions.
The work of MG is supported by the UK Science and Technology Facilities Council (STFC) under Consolidated Grant ST/T000988/1.
SJ acknowledges support by the UK STFC under Consolidated Grant ST/T00102X/1. EvdM acknowledges support through
a PhD studentship funded jointly by STFC/DISCnet and the University of Sussex.

\paragraph{Note added:} We also would like to thank Peter Boyle, Mattia Bruno, Christoph Lehner and Julian Parrino for a comment regarding the projectors, that helped us clarify the definition of the index contractions in Eq.~\eqref{amputated green's function} and Eq.~\eqref{eq:RIcond4}.

\addcontentsline{toc}{section}{\bibname}

\bibliographystyle{jhep}
\bibliography{qed_qcd_matching}
\end{document}